# From Fischer projections to quantum mechanics of tetrahedral molecules: new perspectives in chirality

Salvatore Capozziello[a,*] and Alessandra Lattanzi [b,*]

[a]Dipartimento di Fisica, "E. R. Caianiello" and INFN sez. di Napoli, Università di Salerno, Via S. Allende, 84081 Baronissi, Salerno, Italy

[b]Dipartimento di Chimica, Università di Salerno, Via S. Allende, 84081 Baronissi, Salerno, Italy

**ABSTRACT:** The algebraic structure of central molecular chirality can be achieved starting from the geometrical representation of bonds of tetrahedral molecules, as complex numbers in polar form, and the empirical Fischer projections used in organic chemistry. A general orthogonal $O(4)$ algebra is derived from which we obtain a chirality index $\chi$, related to the classification of a molecule as achiral, diastereoisomer or enantiomer. Consequently, the chiral features of tetrahedral chains can be predicted by means of a molecular Aufbau. Moreover, a consistent Schrödinger equation is developed, whose solutions are the bonds of tetrahedral molecules in complex number representation. Starting from this result, the $O(4)$ algebra can be considered as a "quantum chiral algebra". It is shown that the operators of such an algebra preserve the parity of the whole system.

**CONTENTS**

[*]Correspondence to: A. Lattanzi, Dipartimento di Chimica, Università di Salerno, Via S. Allende, 84081 Baronissi, Salerno, Italy.
E-mails: capozziello@sa.infn.it; lattanzi@unisa.it

## 1. INTRODUCTION

Understanding the fundamental nature of chirality is an issue involving several disciplines of science like physics, chemistry and mathematics.[1] Chirality is a symmetry emerging in abstract spaces (e.g. the spin configuration space of elementary particles) and in real physical space-time (e.g. the Lorentz group of transformations acting on molecules). In any case, it is a discrete symmetry and, from several viewpoints, scientists are wondering if it is a conserved quantity or if it can be violated.[2] In other words, it is not completely clear, up to now, if objects with different states of chirality are different objects or are indistinguishable from a physical point of view (i.e. they have exactly the same energy configuration, the same angular momentum, in modulus, and so on). This question involves subatomic particles (e.g. right-handed and left-handed neutrinos) or molecules (e.g. enantiomers), or even huge macrosystems of astrophysical size as spiral galaxies.[3]

Another and deeper issue is related to the observational fact that Nature seems to prefer, in most cases, just one modality: left-handed neutrinos,[4] L-aminoacids and D-sugars,[5] spiral galaxies with trailing arms.[6] In some sense, we observe a sort of chirality selection rule in our Universe, even if the opposite chiral state is mathematically and physically consistent and can be obtained as a reaction product (e.g. racemization processes and asymmetric synthesis[7]) or happens as secondary process (spiral galaxies with leading arms seem to be the results of interactions in clusters of galaxies).[6]

For example, the origin of homochirality for L-aminoacids and D-sugars is still an open problem and several mechanisms have been proposed.[8] Among them, the spontaneous chiral symmetry breaking represents a fascinating theory. Significant enantiomeric excesses and chirally symmetry breaking can be generated by chirally autocatalytic systems. In particular, the chiral asymmetry generated during a stirred cristallization[9] shows that significant chiral autocatalysis can occur in the proximity of a chiral solid surface; this process might be important to be considered to explain the observed enantiomeric excess in L-amino acids in meteorites.[10]

The program to understand dynamics of chiral structures ranges from microscopic to astrophysical scales, and it is very likely that the whole observable Universe has its own state of chirality.[2a,11]

More specifically, following Lord Kelvin,[12] chirality can be defined as: "I call any geometrical figure, or groups of points, chiral, and say it has chirality, if its image in a plane mirror, ideally realized, cannot be brought to coincide with itself". On the other hand, an achiral molecule can be defined as: "If a structure and its mirror image are superimposable by rotation or any motion other than bond making and breaking, than they are identical". Chiral molecules having central chirality contain stereogenic centres (Fig.1).[13] Given two molecules with identical chemical formulas, if they are not superimposable, they are called *enantiomers*. In general, the term chirality has a broader sense, for example, chirality can be due to a spatial isomerism resulting from the lack of free rotation around single or double bonds (which means that the molecule has a chiral axis) such as in biphenyl[14] compounds (Fig.1).

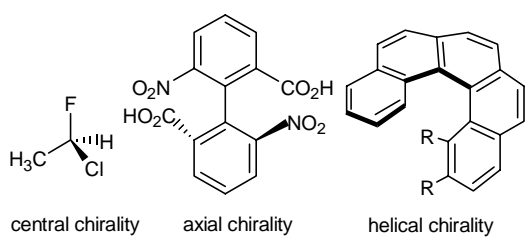

central chirality    axial chirality    helical chirality

**Figure 1.** Different forms of molecular chirality.

Chirality can be even due to a helical shape of the molecule which can be left- or right-handed[15] (Fig.1).

When a molecule contains more than one chiral centre, a further definition has to be introduced. In this case, two molecules with identical structural formulas, which are not mirror images of each other and not superimposable, are termed *diastereoisomers* (Fig.2).

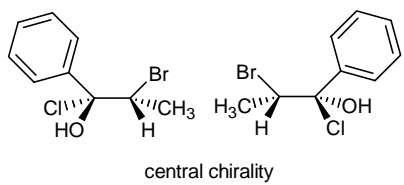

central chirality

**Figure 2.** Example of a couple of diastereoisomers.

Most properties of molecules are invariant under reflection (scalar properties), when examined in an achiral environment and enantiomers are identical in many respects such as solubility, density, melting point, chromatographic retention times, spectroscopic behaviour. It is only with respect to those properties that change sign, but not magnitude under reflection (pseudoscalar properties), that enantiomers differ, such as optical rotation,[16] optical rotatory dispersion (ORD),[17] circular dichroism (CD),[18] vibrational circular dichroism (VCD).[19] In contrast, diastereoisomers exhibit different chemical and physical properties. It is evident that molecular chirality is fundamentally connected to spatial symmetry operations[20] and has the features of a geometrical property. Interestingly, chirality has been treated as a continuous phenomenon[21] of achiral symmetry breaking and this approach has brought to the description of molecules as "more or less chiral" just as a door which is more or less open.

In the last decades, discrete mathematics and qualitative descriptions of the spatial features of molecules provided a large development of theoretical stereochemistry.[22] Molecular chirality has been studied by algebraic methods based on permutation group theory and group representation theory.[23] Several topological indexes have been proposed to describe 3D molecular structures and shapes.[24] Chirality of molecules has been as well the subject of studies aimed to achieve numerical indexes as a measurament of this property, so, discrete and continuous measurements of chirality have been proposed in order to determine the degree of chirality of a molecule.[25] Such measurements are related to the methods used for the characterization of physical and chemical features of a given compound. Empirical classification of organic molecules is based on the properties of functional groups, such as hydroxy group in alcohols, C─C double and triple bonds, CO group in ketones, aldehydes, etc. In general, organic compounds are collected in homologous series, differing by the number of carbons present in the structure. The most important classification of organic molecules, from this point of view, is the Beilstein system, where each compound finds an indexed place in the Beilstein Handbook of Organic Chemistry.[26] Various parameters as thermodynamic enthalpies of formation have been used as a basis for classification in homologous series,[27] and additivity schemes for atoms are introduced in order to predict the enthalpies for compounds of homologous series. Furthermore, quantum mechanical quantities such as molecular total and partial energies have been statistically treated for the same purpose.[28]

The above discussion tells us that several approaches can be pursued but, up to now, scientific community is far from a comprehensive and final theory of chirality. A new approach to figure out the problem could be to plan, as in other fields of science, a sort of "Erlangen program". In fact, according to Felix Klein,[29] every geometry and dinamics of objects can be characterized by their own group of transformations, thus we have, in general the following process

$$\text{Geometry} \Rightarrow \text{Space and Group Transformations} \Rightarrow \text{Dynamics}$$

Following these steps, we can fully characterize a theory, which finally, is well estabilished if experimental data fit the solutions of dynamics.

All physical theories agree with this scheme. As examples, we have

$$\text{Euclidean Space} \Rightarrow \text{Galilei Group} \Leftrightarrow \text{Classical Mechanics}$$

$$\text{Minkowski Space} \Rightarrow \text{Poincaré Group} \Leftrightarrow \text{Special Relativity}$$

$$\text{Phase Space} \Rightarrow \text{Canonical Transformations} \Leftrightarrow \text{Hamiltonian Dynamics}$$

$$\text{Hilbert Space} \Rightarrow \text{Unitary Transformations} \Leftrightarrow \text{Quantum Mechanics}$$

and this approach holds for any self-consistent theory.

Chirality could be dealt, from a theoretical viewpoint, with the same standard by the following steps: 1) given a class of chiral objects identify their configuration space; 2) try to develop the algebra and the group of their configurations and transformations; 3) identify symmetries, conservation laws and then define the dynamical problem; 4) achieve a full theory where objects and their motions are treated at a fundamental level. Following the above schemes, we should have:

$$\text{Configuration Space} \Rightarrow \text{Chirality Transformations} \Leftrightarrow \text{Chiral Mechanics}$$

The approach could involve microscopic (e.g. molecules) and macroscopic (e.g. spiral galaxies) objects, so a full theory of chirality should be a quantum one, but it should work out, in the limits of classical mechanics, also with extremely large objects. In other words, a comprehensive theory of chirality should be independent of objects size.

In this article, this program is outlined for tetrahedral molecules, starting from an elementary geometrical representation.

A tetrahedral molecule is a system of four bonds connected at the origin to a central atom (e.g. a carbon atom). These bonds can be represented as complex numbers in polar form. A chiral transformation between a couple of bonds is nothing else but a complex conjugation and then, taking into account all possible transformations, the elements of the group can be derived. It is interesting to observe that the 24 Fischer projections, coinciding with these elements, constitute an $O(4)$ algebra, so that transformations can be read as rotations and inversions in an abstract 4D-space.

Being the bonds non-relativistic quantum objects, they have to satisfy a Schrödinger equation, so $O(4)$, as it will be shown, can be read as a "quantum chiral algebra" by which it is possible to classify chiral transformations and to construct, in principle, any tetrahedral chain knowing their chiral features.

The layout of the paper is the following. Firstly, the geometry of tetrahedral chains, based on a complex numbers representation, is described. This approach allows a first qualitative classification of molecules as achiral, diastereoisomers and enantiomers. Secondly, we discuss the Fischer projections and show that they can be seen as elements of $O(4)$ algebra. The further step is the extension to a sort of molecular Aufbau for tetrahedral chains. A quantum mechanical approach for tetrahedral molecules is developed by seeking for a consistent Schrödinger equation for bonds. Finally, the quantum chiral algebra is discussed with respect to point transformations and Hund result[30] $\left[\hat{P}, \hat{\mathcal{H}}\right]=0$ is recovered, if parity states are nothing else but superpositions of chiral states.

## 2. GEOMETRICAL APPROACH TO CENTRAL MOLECULAR CHIRALITY BASED ON COMPLEX NUMBERS

The spatial properties of achiral molecules, enantiomers and diastereoisomers can be considered under a geometrical description. Some features exist in order to characterize such classes of molecules by the same parameters.

The approach proposed[31] is based on complex numbers since this is a straightforward way to represent the "length" of the bond with respect to the stereogenic centre and the "angular position" with respect to the other bonds. In general, given a tetrahedral molecule with a stereogenic centre, it can always be projected on a plane containing the stereogenic centre as in Fig.3. Every bond, in the plane {x,y}, can be given in polar representation by

$$\Psi_j = \rho_j e^{i\theta_j} \qquad (1)$$

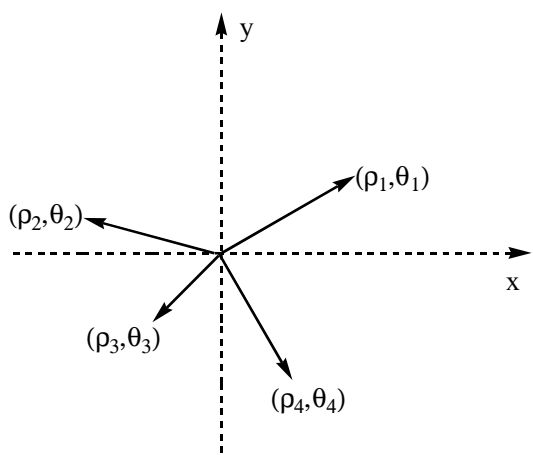

**Figure 3.** Projection of a tetrahedral molecule on a plane containing the stereogenic centre.

where $\rho_j$ is the "modulus", i.e. the projected length of the bond, $\theta_j$ is the "anomaly", i.e. the position of the bond with respect to the x, y axes (and then with respect to the other bonds). The number $i = \sqrt{-1}$ is the imaginary unit. A molecule with one stereogenic centre is then given by the sum vector

$$\mathcal{M} = \sum_{j=1}^{4} \rho_j e^{i\theta_j} \qquad (2)$$

in any symmetry plane. If the molecule has *n* stereogenic centres, we can define *n* planes of projection (one for each centre). Such planes can be parallel among them, even if this feature is not essential. If a molecule with one centre has four bonds, a molecule with two centres has seven bonds and so on. The general rule is

$$n = \text{centres} \quad \Leftrightarrow \quad 4n - (n-1) = 3n+1 \text{ bonds} \qquad (3)$$

assuming simply connected tetrahedra. If atoms acting as "spacers" are present between chiral centres, the number of bonds changes from $3n+1$ to $4n$, but the following considerations for consecutive connected tetrahedra remain valid. A molecule with *n* stereogenic centres is then given by the sum vector

$$\mathcal{M}_n = \sum_{k=1}^{n} \sum_{j=1}^{3n+1} \rho_{jk} e^{i\theta_{jk}} \qquad (4)$$

where k is the "centre-index" and j is the "bond-index". Again, for any k, a projective plane of symmetry is defined. The couple of numbers $\{\rho, \theta\} \equiv \{0,0\}$ assigns the centre in every plane. In other words, a molecule $\mathcal{M}_n$ is assigned by the two sets of numbers

$$\{\rho_{1k}, \ldots \rho_{jk}, \ldots \rho_{(3n+1)k}\}$$

$$\{\theta_{1k}, \ldots \theta_{jk}, \ldots \theta_{(3n+1)k}\} \qquad (5)$$

Having in mind the definition of chirality, the behaviour of the molecule under rotation and superimposition has to be studied in order to check if the structure and its mirror image are superimposable. Chirality emerges when two molecules with identical structural formulas are not superimposable. Considering the geometrical representation reported in Fig.3, a possible situation is the following: let us take into account a rotation of 180° in the space around a generic axis $L$ passing through the stereogenic centre. Such an axis can coincide, for the sake of simplicity, with one of the bonds. After the rotation, two bonds result surimposable while the other two are inverted. The situation can be illustrated by the projection on the plane {x,y} as shown in Fig.4.

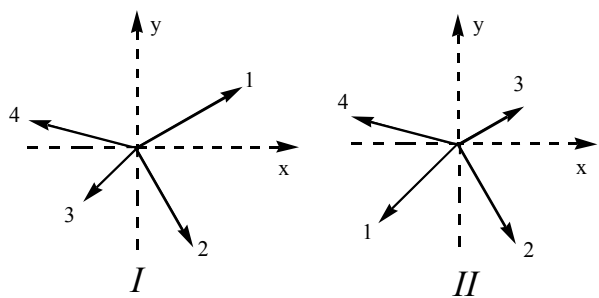

**Figure 4.** Picture of the projected situations before and after the rotation of a chiral tetrahedron over its mirror image. Groups 2 and 4 coincide while 3 and 1 are inverted.

In formulae, for the inverted bonds, we have

$$\{\Psi_1 = \rho_1 e^{i\theta_1}, \Psi_3 = \rho_3 e^{i\theta_3}\} \Rightarrow \{\overline{\Psi}_1 = \rho_1 e^{i\theta_3}, \overline{\Psi}_3 = \rho_3 e^{i\theta_1}\} \qquad (6)$$

In order to observe the reflection, the four groups must be of different nature. This simple observation shows that the chirality is connected with an inversion of two bonds in the projective symmetry plane. On the contrary, if after the rotation and superimposition (Fig.4), molecule *I* is identical to molecule *II*, we are in an achiral situation. Such a treatment can be repeated for any projective symmetry plane which can be defined for the $n$ centres. The possible results are that the molecule is fully invariant after rotation(s) and superimposition with respect to its mirror image (achiral); the molecule is partially invariant after rotation(s) and superimposition, i.e. some tetrahedra are superimposable while others are not (diastereoisomers); the molecule presents an inversion for each stereogenic centre (enantiomers).

The following rule can be derived: central chirality is assigned by the number $\chi$ given by the couple $n, p$ that is

$$\chi = \{n, p\} \tag{7}$$

where $\chi$ is the chirality index, $n$ is the principal chiral number and $p$ the secondary chiral number, that is $n$ is the number of stereogenic centres, $p$ is the number of permutations (at most one for any centre). The constraint

$$0 \leq p \leq n \tag{8}$$

has to hold.

This definition of chirality is related to the structure of the molecule and its properties under rotations and superimposition.

The sequence between achiral and chiral molecules is given by

$$\chi \equiv \{n, 0\} \quad \text{achiral molecules}$$

$$\chi \equiv \{n, p < n\} \quad \text{diastereoisomers}$$

$$\chi \equiv \{n, n\} \quad \text{enantiomers}$$

As an example, the chirality of degenerate case, *meso*-tartaric acid (Fig.5), can be reduced to this rule. In this case, three groups of two tetrahedra are identical and the fourth group is the other stereogenic carbon centre. As it can be seen from the figure, the molecule is fully invariant by superimposition to its mirror image, hence $p = 0$ and the structure is achiral ($\chi \equiv \{2, 0\}$).

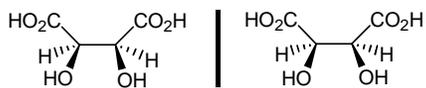

**Figure 5.** Mirror structures of *meso*-tartaric acid.

## 3. FISCHER PROJECTIONS FOR TETRAHEDRAL MOLECULES

An extremely useful method to represent tetrahedral molecules was reported in 1891 by Emil Fischer, who proposed the well-known planar projection formulas. When describing a molecule in this representation, some rules have to be followed (Fig.6): the atoms pointing sideways must project forward in the model, while those pointing up and down in the projection must extend toward the rear. As an example, let us take into account (*S*)-(+)-lactic acid.

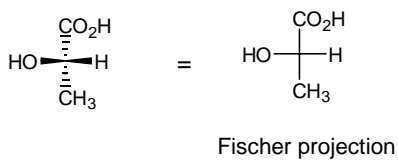

Fischer projection

**Figure 6.** Fischer projection of (*S*)-(+)-lactic acid.

In order to obtain proper results using Fischer projections, they must be treated differently from models in testing superimposability. Projections may not be rotated of 90°, while a 180° rotation is allowed. The interchange of any two groups results in the conversion of an enantiomer into its mirror image (Fig.7).

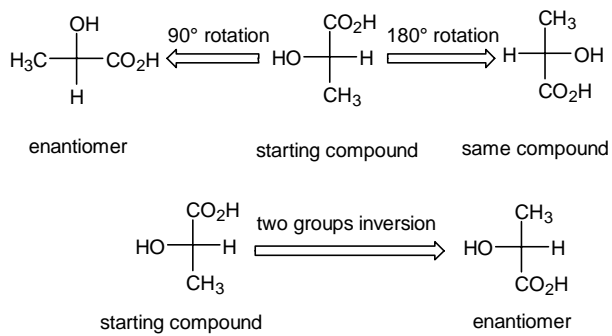

**Figure 7.** Fundamental rules to handle Fischer projections.

Let us indicate the chemical groups by numbers running from 1 to 4. For the example which we are considering: OH=1, $CO_2H$=2, H=3, $CH_3$=4, without taking into account the effective priorities of the groups.[13] There are 24 (=4! the number of permutations of 4 ligands among 4 sites) projections.

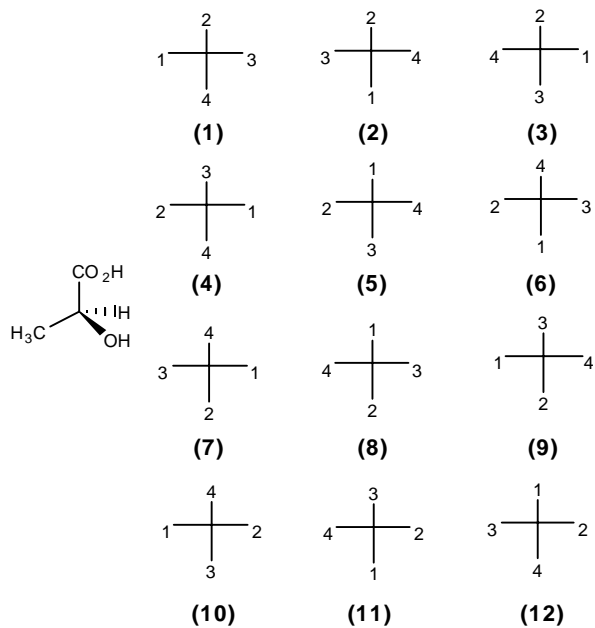

**Figure 8.** Twelve Fischer projections of (S)-(+)-lactic acid.

Twelve of these correspond to the (+) enantiomer and are illustrated in Fig.8.

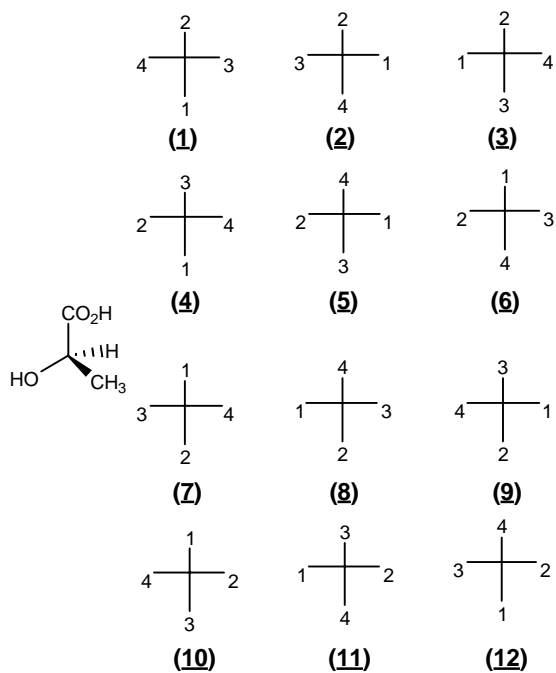

**Figure 9.** Twelve Fischer projections of (*R*)-(−)-lactic acid.

The other 12 graphs in Fig. 9 represent the (−) enantiomer.

The permutations shown in Fig.8 can be obtained, either by permuting groups of three bonds or by turning the projections by 180°. The permutations outlined in Fig.9 derive by those in Fig. 8 simply by interchanging two groups. With these considerations in mind, it is immediate asking for an algebraic structure which can be built from Fischer projections.[32]

## 4. ALGEBRAIC STRUCTURE OF CENTRAL MOLECULAR CHIRALITY

In order to reduce the Fischer rules to an algebraic structure, we define an operator $\chi_k$ acting on a tetrahedral molecule. We shall take into account only one tetrahedron, but the generalization of the following results to simply connected chains of tetrahedra is easily accomplished as we shall see below.

A tetrahedral molecule can be assigned by a column vector $\mathcal{M}$, rewriting Eq. (2) as

$$\mathcal{M} = \begin{pmatrix} \Psi_1 \\ \Psi_2 \\ \Psi_3 \\ \Psi_4 \end{pmatrix} \qquad (9)$$

$\Psi_j$ are defined in Eq. (1).

The corresponding Fischer projection is

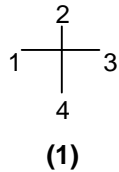

**(1)**

which is the first in Fig.8. The position of the bonds in the column vector (9) are assigned starting from the left and proceeding clockwise in the Fischer projection.

The matrix representation of the projection **(1)** is assumed as "fundamental", i.e.

$$\chi_1 = \begin{pmatrix} 1 & 0 & 0 & 0 \\ 0 & 1 & 0 & 0 \\ 0 & 0 & 1 & 0 \\ 0 & 0 & 0 & 1 \end{pmatrix} \qquad (10)$$

so the action on the column vector $\mathcal{M}$ is

$$\chi_1 \begin{pmatrix} \Psi_1 \\ \Psi_2 \\ \Psi_3 \\ \Psi_4 \end{pmatrix} = \begin{pmatrix} \Psi_1 \\ \Psi_2 \\ \Psi_3 \\ \Psi_4 \end{pmatrix} \qquad (11)$$

$\chi_1$ is nothing else but the identity operator. The configuration **(2)** of Fig.8 can be achieved as soon as we define an operator $\chi_2$ acting as

$$\chi_2 \begin{pmatrix} \Psi_1 \\ \Psi_2 \\ \Psi_3 \\ \Psi_4 \end{pmatrix} = \begin{pmatrix} \Psi_3 \\ \Psi_2 \\ \Psi_4 \\ \Psi_1 \end{pmatrix} \tag{12}$$

which corresponds to the matrix

$$\chi_2 = \begin{pmatrix} 0 & 0 & 1 & 0 \\ 0 & 1 & 0 & 0 \\ 0 & 0 & 0 & 1 \\ 1 & 0 & 0 & 0 \end{pmatrix} \tag{13}$$

it is clear that $\chi_2$ is a rotation. On the other hand, the configuration **(1)** of the (−) enantiomer can be obtained starting from the column vector (9), if we define an operator $\bar{\chi}_1$ which acts as

$$\bar{\chi}_1 \begin{pmatrix} \Psi_1 \\ \Psi_2 \\ \Psi_3 \\ \Psi_4 \end{pmatrix} = \begin{pmatrix} \Psi_4 \\ \Psi_2 \\ \Psi_3 \\ \Psi_1 \end{pmatrix} \tag{14}$$

Explicitly, we have

$$\bar{\chi}_1 = \begin{pmatrix} 0 & 0 & 0 & 1 \\ 0 & 1 & 0 & 0 \\ 0 & 0 & 1 & 0 \\ 1 & 0 & 0 & 0 \end{pmatrix} \tag{15}$$

It generates the inversion between the bonds $\Psi_1$ and $\Psi_4$.

By this approach, all the 24 projections can be obtained (12 for the (+) enantiomer and 12 for the (−) enantiomer represented in Figs.8 and 9) by matrix operators acting on the fundamental projection **(1)**. The following tables summarize the situation. The operators $\chi_k$ give rise to the representations of the (+) enantiomer, while the operators $\bar{\chi}_k$ give rise to those of the (−) enantiomer. Obviously $k = 1,..,12$.

Table I, (+)-enantiomer:

$$\chi_1 = \begin{pmatrix} 1 & 0 & 0 & 0 \\ 0 & 1 & 0 & 0 \\ 0 & 0 & 1 & 0 \\ 0 & 0 & 0 & 1 \end{pmatrix} \quad \chi_2 = \begin{pmatrix} 0 & 0 & 1 & 0 \\ 0 & 1 & 0 & 0 \\ 0 & 0 & 0 & 1 \\ 1 & 0 & 0 & 0 \end{pmatrix} \quad \chi_3 = \begin{pmatrix} 0 & 0 & 0 & 1 \\ 0 & 1 & 0 & 0 \\ 1 & 0 & 0 & 0 \\ 0 & 0 & 1 & 0 \end{pmatrix}$$

$$\chi_4 = \begin{pmatrix} 0 & 1 & 0 & 0 \\ 0 & 0 & 1 & 0 \\ 1 & 0 & 0 & 0 \\ 0 & 0 & 0 & 1 \end{pmatrix} \quad \chi_5 = \begin{pmatrix} 0 & 1 & 0 & 0 \\ 1 & 0 & 0 & 0 \\ 0 & 0 & 0 & 1 \\ 0 & 0 & 1 & 0 \end{pmatrix} \quad \chi_6 = \begin{pmatrix} 0 & 1 & 0 & 0 \\ 0 & 0 & 0 & 1 \\ 0 & 0 & 1 & 0 \\ 1 & 0 & 0 & 0 \end{pmatrix}$$

$$\chi_7 = \begin{pmatrix} 0 & 0 & 1 & 0 \\ 0 & 0 & 0 & 1 \\ 1 & 0 & 0 & 0 \\ 0 & 1 & 0 & 0 \end{pmatrix} \quad \chi_8 = \begin{pmatrix} 0 & 0 & 0 & 1 \\ 1 & 0 & 0 & 0 \\ 0 & 0 & 1 & 0 \\ 0 & 1 & 0 & 0 \end{pmatrix} \quad \chi_9 = \begin{pmatrix} 1 & 0 & 0 & 0 \\ 0 & 0 & 1 & 0 \\ 0 & 0 & 0 & 1 \\ 0 & 1 & 0 & 0 \end{pmatrix}$$

$$\chi_{10} = \begin{pmatrix} 1 & 0 & 0 & 0 \\ 0 & 0 & 0 & 1 \\ 0 & 1 & 0 & 0 \\ 0 & 0 & 1 & 0 \end{pmatrix} \quad \chi_{11} = \begin{pmatrix} 0 & 0 & 0 & 1 \\ 0 & 0 & 1 & 0 \\ 0 & 1 & 0 & 0 \\ 1 & 0 & 0 & 0 \end{pmatrix} \quad \chi_{12} = \begin{pmatrix} 0 & 0 & 1 & 0 \\ 1 & 0 & 0 & 0 \\ 0 & 1 & 0 & 0 \\ 0 & 0 & 0 & 1 \end{pmatrix}$$

Table II, (−) enantiomer:

$$\bar{\chi}_1 = \begin{pmatrix} 0 & 0 & 0 & 1 \\ 0 & 1 & 0 & 0 \\ 0 & 0 & 1 & 0 \\ 1 & 0 & 0 & 0 \end{pmatrix} \quad \bar{\chi}_2 = \begin{pmatrix} 0 & 0 & 1 & 0 \\ 0 & 1 & 0 & 0 \\ 1 & 0 & 0 & 0 \\ 0 & 0 & 0 & 1 \end{pmatrix} \quad \bar{\chi}_3 = \begin{pmatrix} 1 & 0 & 0 & 0 \\ 0 & 1 & 0 & 0 \\ 0 & 0 & 0 & 1 \\ 0 & 0 & 1 & 0 \end{pmatrix}$$

$$\bar{\chi}_4 = \begin{pmatrix} 0 & 1 & 0 & 0 \\ 0 & 0 & 1 & 0 \\ 0 & 0 & 0 & 1 \\ 1 & 0 & 0 & 0 \end{pmatrix} \quad \bar{\chi}_5 = \begin{pmatrix} 0 & 1 & 0 & 0 \\ 0 & 0 & 0 & 1 \\ 1 & 0 & 0 & 0 \\ 0 & 0 & 1 & 0 \end{pmatrix} \quad \bar{\chi}_6 = \begin{pmatrix} 0 & 1 & 0 & 0 \\ 1 & 0 & 0 & 0 \\ 0 & 0 & 1 & 0 \\ 0 & 0 & 0 & 1 \end{pmatrix}$$

$$\overline{\chi}_7 = \begin{pmatrix} 0 & 0 & 1 & 0 \\ 1 & 0 & 0 & 0 \\ 0 & 0 & 0 & 1 \\ 0 & 1 & 0 & 0 \end{pmatrix} \quad \overline{\chi}_8 = \begin{pmatrix} 1 & 0 & 0 & 0 \\ 0 & 0 & 0 & 1 \\ 0 & 0 & 1 & 0 \\ 0 & 1 & 0 & 0 \end{pmatrix} \quad \overline{\chi}_9 = \begin{pmatrix} 0 & 0 & 0 & 1 \\ 0 & 0 & 1 & 0 \\ 1 & 0 & 0 & 0 \\ 0 & 1 & 0 & 0 \end{pmatrix}$$

$$\overline{\chi}_{10} = \begin{pmatrix} 0 & 0 & 0 & 1 \\ 1 & 0 & 0 & 0 \\ 0 & 1 & 0 & 0 \\ 0 & 0 & 1 & 0 \end{pmatrix} \quad \overline{\chi}_{11} = \begin{pmatrix} 1 & 0 & 0 & 0 \\ 0 & 0 & 1 & 0 \\ 0 & 1 & 0 & 0 \\ 0 & 0 & 0 & 1 \end{pmatrix} \quad \overline{\chi}_{12} = \begin{pmatrix} 0 & 0 & 1 & 0 \\ 0 & 0 & 0 & 1 \\ 0 & 1 & 0 & 0 \\ 1 & 0 & 0 & 0 \end{pmatrix}$$

The matrices in Table I and II are the elements of a 4-parameter algebra. Those in Table I are a representation of rotations, while those in Table II are inversions. Both sets constitute the group $O(4)$ of $4 \times 4$ orthogonal matrices. The matrices in Table I are the remarkable subgroup $SO(4)$ of $4 \times 4$ matrices with determinant $+1$. The matrices in Table II have determinant $-1$, being inversions (or reflections). They do not constitute a group since the product of any two of them has determinant $+1$. This fact means that the product of two inversions generates a rotation (this is obvious by inverting both the couples of bonds in a tetrahedron). In fact, we have

$$\chi_k \chi_l = \chi_m, \quad \overline{\chi}_k \overline{\chi}_l = \chi_m, \quad \overline{\chi}_k \chi_l = \overline{\chi}_m \quad \text{for} \quad k,l,m,=1,...,12 \qquad (16)$$

For example, straightforward calculations give

$$\chi_8 \chi_9 = \chi_5, \quad \overline{\chi}_5 \overline{\chi}_2 = \chi_9, \quad \overline{\chi}_{10} \chi_{10} = \overline{\chi}_7 \qquad (17)$$

and so on. In summary, the product of two rotations is a rotation, the product of two reflections is a rotation, while the product of a reflection and a rotation is again a reflection. In any case, the total algebra is closed.[33] Below the complete set of commutation relations is given.

The 24 matrices in Table I and II are not all independent. They can be grouped as different representations of the same operators. To this aim, we make use of a fundamental theorem of algebra which states that all matrices, representing the same operator, have the same characteristic polynomial.[34] In other words, the characteristic equation of a matrix is invariant under vector base changes. In (+) enantiomer case, the characteristic eigenvalue equation is

$$\det\|\chi_k - \lambda \mathbf{I}\| = 0 \tag{18}$$

where $\lambda$ are the eigenvalues and $\mathbf{I}$ is the identity matrix.

The following characteristic polynomials can be derived

$$(1-\lambda)^4 = 0 \qquad \text{for } \chi_1 \tag{19}$$

$$(1-\lambda)^2(1+\lambda+\lambda^2) = 0 \qquad \text{for } \chi_2, \chi_3, \chi_4, \chi_6, \chi_8, \chi_9, \chi_{10}, \chi_{12} \tag{20}$$

$$(1-\lambda)^2(1+\lambda)^2 = 0 \qquad \text{for } \chi_5, \chi_7, \chi_{11} \tag{21}$$

In the case of (−) enantiomer, we have

$$\det\|\bar{\chi}_k - \lambda \mathbf{I}\| = 0 \tag{22}$$

and the characteristic polynomials are

$$(1-\lambda)^3(1+\lambda) = 0 \qquad \text{for } \bar{\chi}_1, \bar{\chi}_2, \bar{\chi}_3, \bar{\chi}_6, \bar{\chi}_8, \bar{\chi}_{11} \tag{23}$$

$$(1-\lambda)(1+\lambda)(\lambda^2+1) = 0 \quad \text{for } \bar{\chi}_4, \bar{\chi}_5, \bar{\chi}_7, \bar{\chi}_9, \bar{\chi}_{10}, \bar{\chi}_{12} \tag{24}$$

There are 6 independent eigenvalues:

$$\lambda_{1,2} = \pm 1 \qquad \lambda_{3,4} = \pm i \qquad \lambda_{5,6} = \frac{-1 \pm i\sqrt{3}}{2} \tag{25}$$

Inserting them into Eqs. (18) and (22), it is easy to determine the eigenvectors

$$(\chi_k - \lambda \mathbf{I})\mathcal{M} = 0, \qquad (\bar{\chi}_k - \lambda \mathbf{I})\mathcal{M} = 0 \tag{26}$$

with obvious calculations depending on the choice of $\chi_k$ and $\bar{\chi}_k$. $\mathcal{M}$ is given by Eq. (9). It is worth noting that the number of independent eigenvalues (and then eigenvectors) is related to the number of independent elements in each member of the group $O(N)$ we are considering. $N^2$ is the total number of elements, while $\frac{1}{2}N(N+1)$ are the orthogonality conditions, so we have

$$N^2 - \frac{1}{2}N(N+1) = \frac{1}{2}N(N-1) \tag{27}$$

For $O(4)$, it is 6, which is the number of independent generators of the group,[33] giving the "dimension" of the group. With these considerations in mind, it can be stated that Fischer projections generates the algebraic structure of tetrahedral molecules.

## 5. GENERALIZATION TO MOLECULES WITH N STEREOGENIC CENTRES: A MOLECULAR AUFBAU FOR TETRAHEDRAL CHAINS

The results of the previous section can be extended to more general cases. For a molecule with $n$ stereogenic centres, we can define $n$ planes of projection and the bonds among the centres have to be taken into account.

Eq. (4) can be written as

$$\mathcal{M}_n = \sum_{k=1}^{p} \overline{\mathcal{M}}_k + \sum_{k=p+1}^{n} \mathcal{M}_k \tag{28}$$

where $\overline{\mathcal{M}}_k$ and $\mathcal{M}_k$ are generic tetrahedra on which are acting the operators $\overline{\chi}_l^k$ and $\chi_l^k$ respectively; k is the center index running from 1 to $n$; l is the operator index ranging from 1 to 12.

For any tetrahedron, two possibilities are available:

$$\mathcal{M}_k = \chi_l^k \mathcal{M}_k^{(0)}, \qquad \overline{\mathcal{M}}_k = \overline{\chi}_l^k \mathcal{M}_k^{(0)} \tag{29}$$

where $\mathcal{M}_k^{(0)}$ is the starting fundamental representation of the k-tetrahedron given by the column vector in Eq. (9). Explicitly we have

$$\mathcal{M}_k^{(0)} = \begin{pmatrix} \Psi_{1k} \\ \Psi_{2k} \\ \Psi_{3k} \\ \Psi_{4k} \end{pmatrix} \qquad (30)$$

In other words, $\mathcal{M}_k$ and $\overline{\mathcal{M}}_k$ are the result of the application of one of the above matrix operators on the starting column vector $\mathcal{M}_k^{(0)}$.

A particular discussion deserves the index $p$, which, as previously stated, ranges $0 \leq p \leq n$. It is the number of permutations, which occur when the operators $\overline{\chi}_1^k$ act on tetrahedra. It corresponds to the number of reflections occuring in a $n$-centre tetrahedral chain. No inversions, but rotations occur when $\chi_1^k$ operators act on the molecule. Having this rule in mind, it follows that

$$\mathcal{M}_n = \sum_{k=1}^{n} \mathcal{M}_k, \qquad p = 0 \qquad (31)$$

is an achiral molecule (in this case only $\chi_1^k$ operators act on $\mathcal{M}_k^{(0)}$);

$$\mathcal{M}_n = \sum_{k=1}^{p} \overline{\mathcal{M}}_k + \sum_{k=p+1}^{n} \mathcal{M}_k, \qquad 0 < p < n \qquad (32)$$

is a diastereoisomer since $[n-(p+1)]$ tetrahedra result superimposable after rotations, while $p$-ones are not superimposable, having, each of them, undergone an inversion of two of their bonds.

Finally, an enantiomer results if

$$\mathcal{M}_n = \sum_{k=1}^{n} \overline{\mathcal{M}}_k, \qquad n = p \qquad (33)$$

where every tetrahedron results a mirror image of its starting situation after the application of any of the $\overline{\chi}_1^k$ operators. The chirality selection rule, geometrically deduced [Eq. (7)], is fully recovered. In other words, this selection rule gives a classification of tetrahedral chains by their chirality structure.

A building-up process (Aufbau-like) is consequently derivable.[35] The building-up process gives rise to a chirality index which assigns the intrinsic chiral structure of the final compound. As we have seen, the chirality index $\chi$ allows an immediate chiral characterization of a given tetrahedral chain. Let us take into account a molecule, which is well-defined in its chiral feature, in the sense that, considering also its mirror image, it is clear to assess if the molecule is an enantiomer, a diastereoisomer or an achiral molecule. After the addition of a further chiral centre to this structure and its mirror image, the resulting structure will be

$$\chi \equiv \{n+1, p+\Delta p\} \tag{34}$$

where $\Delta p = 0,1$. The chiral properties of the new molecule are assigned by the $\Delta p$ value according to the following possibilities.

If $\Delta p = 0$, we can have

$$\chi_s \equiv \{n,0\} \Rightarrow \chi_f \equiv \{n+1,0\} \tag{35}$$

in this case, the starting compound is an achiral molecule as well as the final one.

Again, for $\Delta p = 0$, we can have

$$\chi_s \equiv \{n, p\} \Rightarrow \chi_f \equiv \{n+1, p\} \tag{36}$$

in this case, the starting molecule is a diastereoisomer, being $n > p$, as well as the final structure.

Finally, if

$$\chi_s \equiv \{n, n\} \Rightarrow \chi_f \equiv \{n+1, n\} \tag{37}$$

the starting molecule is an enantiomer, while the final one is a diastereoisomer.

If $\Delta p = 1$, the situations can be

$$\chi_s \equiv \{n,0\} \Rightarrow \chi_f \equiv \{n+1,1\} \tag{38}$$

from an achiral molecule, a diastereoisomer is obtained;

$$\chi_s \equiv \{n, p\} \Rightarrow \chi_f \equiv \{n+1, p+1\} \qquad (39)$$

from a diastereoisomer, another diastereoisomer is obtained;

$$\chi_s \equiv \{n, n\} \Rightarrow \chi_f \equiv \{n+1, n+1\} \qquad (40)$$

from an enantiomer, we get another enantiomer.

Eqs. (35)-(40) take into account all the possibilities, which can be easily iterated adding up any number of chiral centres to a given chain. In the general case, the *Aufbau rule* is

$$\chi \equiv \{n+n', p+p'\}; \forall n' \geq 1, \quad p' = \sum_{j=1}^{n'} \Delta p_j, \ \Delta p_j = 0,1 \qquad (41)$$

However, we have to consider that the rule works only for simply connected tetrahedral chains, where the chiral features are well-estabilished with respect to the mirror image. In this sense, chirality is not an absolute feature of the molecules.[2a, 20] Adding up a chiral centre to a structure gives rise to a new molecule, where $\chi \equiv \{n+1, p+\Delta p\}$. The fact that, in the addition, the variation of $p$ can be $\Delta p = 0,1$, assigns the chiral feature of the new compound.

An example of the building-up process is reported in Fig.10. In Fig.11, a degenerate case[13] is reported, where the two chiral centres are identical, introducing a further degree of symmetry to the final structure. In this case, the situation $\chi_f \equiv \{2,2\}$ for $\Delta p = 1$ is equivalent to $\chi_f \equiv \{2,0\}$, since the two molecules are superimposable, hence the structure is achiral.

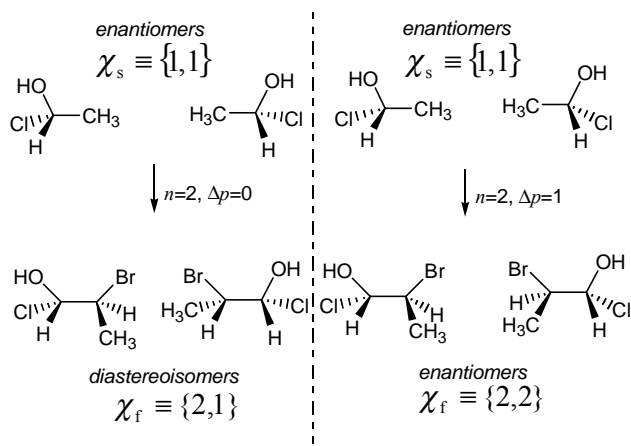

**Figure 10.** Aufbau process consisting in adding up a chiral centre to a given chiral tetrahedron.

Last consideration indicates that such an Aufbau approach is working only if the chiral centres are different and, in this respect, the procedure is suitable for the description of chiral structures.

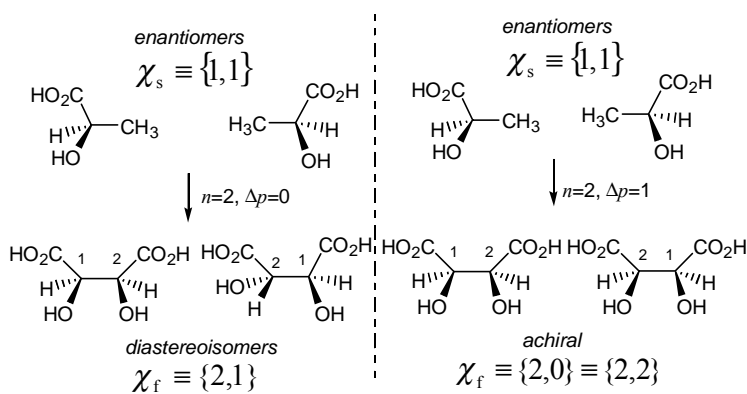

**Figure 11.** A degenerate example of Aufbau process consisting in adding up a chiral centre, having identical substituents of the starting chiral tetrahedron.

## 6. QUANTUM MECHANICAL APPROACH

In order to carry on with our Erlangen-like program, the next step required is dealing with dynamics of tetrahedra starting from the previous geometrical and algebraic considerations.

Being microscopic objects, a quantum mechanical treatment has to be pursued. First of all, we have to probe if the mathematical representation of bond given in Eq. (1), and then the superpositions (2) and (4), are solutions of a suitable Schrödinger equation. Furthermore, in order to build up a self-consistent quantum mechanics of chiral eigenstates, relations among such eigenstates, energy and parity eigenstates need to be found.[20] The final step is to understand which is the fundamental meaning of chirality transformations.

In order to answer such questions, we have to discuss the possible quantum mechanical interpretation of the above formulae and then the consistency of the problem in the perspective of a full quantum mechanical treatment.

We proceed by an inverse problem approach, considering a Schrödinger equation for the "solution" (1) and the sums of solutions (2) and (4).[36]

The problem is set in the Born-Oppenheimer approximation by which a given stereogenic centre is considered fixed and dynamics of the four bonds is reduced to it. This position is supported by an appropriate change of coordinates, since the problem can be reduced to a coordinate system fixed in the stereogenic centre.

A time-independent three-dimensional Schrödinger equation is

$$\left\{-\frac{\hbar^2}{2\mu}\nabla^2 + V(x,y,z)\right\}\eta(x,y,z) = E_j\,\eta(x,y,z) \qquad (42)$$

where $E_j$ are the energy eigenvalues, "$\mu$" is a given reduced mass and $\nabla^2$ the Laplace operator. The index j will be defined below. Our tetrahedron can be idealized as a system in a central potential with a spherical symmetry, so that, the potential depends only on the radius r: $V(\bar{r}) \equiv V(r)$. The general solution of the angular part of Schrödinger equation is

$$Y_{l,m}(\vartheta,\varphi) = N_{l,m} P_l^m(\cos\varphi)(\sin\varphi)^{|m|} e^{im\vartheta} \quad (43)$$

where $P_l^m(\cos\varphi)$ are Legendre polynomials and $N_{l,m}$ is a normalization factor depending on the orbital and azimuthal quantum numbers $l$ and $m$. Coming back to the former problem, the aim was to see if Eq. (1) is a solution of Schrödinger equation. Concerning the angular component, it can be interpreted as the azimuthal part of angular momentum.

For the radial component, it can be chosen

$$\rho(r) = \left(\frac{r}{r_0}\right) \quad (44)$$

where $r_0$ is a normalization length (e.g. $\approx 1.09 \div 1.54$ Å a typical C—X length bond) useful to restore the probabilistic interpretation of our approach. Immediately, the form of the potential $V(r)$ is obtained

$$V(r) = E_j - \frac{\alpha_0}{r^2}, \qquad \alpha_0 = \frac{\hbar^2}{2\mu}[l(l+1)-2] \quad (45)$$

depending on the eigenvalue $E_j$, the angular momentum $l$ and the mass $\mu$.

Finally, the solution $\Psi_j = \rho_j e^{i\vartheta_j}$ with the positions

$$\rho_j(r) = \left(\frac{r_j}{r_0}\right), \qquad \vartheta_j = m_j \vartheta \quad (46)$$

is composed by the radial and azimuthal parts of a complete solution of Schrödinger equation. Clearly, $j=1,2,3,4$, $E_j$ are the bonds energies and $\mathcal{M} = \sum_{j=1}^{4} \rho_j e^{i\vartheta_j}$ is the superposition of four single particular solutions. This result can be extended to Eq. (4) considering $n$ Schrödinger problems, one for each stereogenic centre.[36] At this point, the role of operators $\chi_s^k$ and $\bar{\chi}_s^k$ has to be investigated in order to see if they are quantum operators implementing chiral transformations.

## 7. QUANTUM CHIRAL ALGEBRA AND PARITY

Previous treatment shows that Eq. (1) can be considered as a solution of a "reduced" Schrödinger problem, where, due to the separation of variables, the Hamiltonian operator is projected on the $\{r,\vartheta\}$- plane and a part of the general solution $\Phi = \Phi(r,\vartheta,\varphi,t)$ is nothing else but $\Psi(r,\vartheta) = \rho e^{i\vartheta}$. Operators $\hat{\mathcal{H}}_k(r,\vartheta)$, $\chi_s^k$ and $\bar{\chi}_s^k$, act on the four bonds of the k-stereogenic centre inducing the following transformations

$$i\hbar \frac{\partial}{\partial t}\mathcal{M}_k = \hat{\mathcal{H}}_k \mathcal{M}_k = E_{jk} \mathcal{M}_k, \qquad \mathcal{M}_k = \chi_s^k \mathcal{M}_k^{(0)} \qquad (47)$$

and

$$i\hbar \frac{\partial}{\partial t}\overline{\mathcal{M}}_k = \hat{\mathcal{H}}_k \overline{\mathcal{M}}_k = E_{jk} \overline{\mathcal{M}}_k, \qquad \overline{\mathcal{M}}_k = \bar{\chi}_s^k \mathcal{M}_k^{(0)} \qquad (48)$$

where $E_{jk}$ are the energy egeinstates of bonds with respect to the k-stereogenic centre. Rotations $\chi_s^k$ and inversions $\bar{\chi}_s^k$ operate on bonds of the starting fundamental representation $\mathcal{M}_k^{(0)}$ given by Eq. (30).

It is straightforward to see that $\chi_s^k$ and $\hat{\mathcal{H}}_k$ commute between them being

$$\left[\hat{\mathcal{H}}_k, \chi_s^k\right] = 0 \qquad (49)$$

Similarly, it can be obtained

$$\left[\hat{\mathcal{H}}_k, \bar{\chi}_s^k\right] = 0 \qquad (50)$$

Furthermore, we have

$$\left[\chi_s^k, \chi_m^k\right] = 0 \qquad \text{for } s,m = 5,7,11 \qquad (51)$$

$$\left[\bar{\chi}_s^k, \bar{\chi}_m^k\right] = \chi_l^k - \chi_g^k \quad \text{for } l, m, g, s = 1, ..., 12 \tag{52}$$

$$\left[\chi_s^k, \chi_m^k\right] = \chi_l^k - \chi_g^k \quad \text{for } l, m, g, s = 1, ..., 12 \text{ with } s, m \neq 5, 7, 11 \tag{53}$$

$$\left[\chi_s^k, \bar{\chi}_m^k\right] = \chi_l^k - \chi_g^k \quad \text{for } l, m, g, s = 1, ..., 12 \tag{54}$$

and

$$\left[\chi_s^k, \chi_m^g\right] = 0, \quad \left[\bar{\chi}_s^k, \bar{\chi}_m^g\right] = 0, \quad \left[\bar{\chi}_s^k, \chi_m^g\right] = 0 \tag{55}$$

being $k \neq g$.

Relations (49)-(55) constitute a quantum chiral algebra and the eigenstates of $\chi_s^k$ and $\bar{\chi}_s^k$ operators are, in general, solutions of Schrödinger equation. The way in which operators $\hat{\mathcal{H}}_k$, $\chi_s^k$, $\bar{\chi}_s^k$ work on quantum states of chiral molecules deserves a particular discussion.

Taking into account the fundamental representation (30) of a given k-tetrahedron, the action of the operator $\chi_s^k$ and $\bar{\chi}_s^k$ defines the "chiral state" of the molecule being

$$|\Psi_R\rangle = \mathcal{M}_k = \chi_s^k \mathcal{M}_k^{(0)} \tag{56}$$

and

$$|\Psi_L\rangle = \overline{\mathcal{M}}_k = \bar{\chi}_s^k \mathcal{M}_k^{(0)} \tag{57}$$

where $|\Psi_R\rangle$ and $|\Psi_L\rangle$ indicate right- and left-handed quantum states of the molecule using the Dirac ket notation. Operators $\chi_s^k$ "rotate" the k-tetrahedron, while $\bar{\chi}_s^k$ "invert" a couple of bonds. Dropping, for simplicity, the indexes, the following relations

$$\chi|\Psi_R\rangle = |\Psi_R\rangle; \quad \chi|\Psi_L\rangle = |\Psi_L\rangle \tag{58}$$

$$\bar{\chi}|\Psi_R\rangle = |\Psi_L\rangle; \quad \bar{\chi}|\Psi_L\rangle = |\Psi_R\rangle \tag{59}$$

hold. The $\bar{\chi}$ operators interconvert two handed forms and, in some sense, work as an algebraic counterpart of quantum tunnelling.[2a,37] Parity eigenstates of a chiral molecule, ignoring parity violation effects,[38] are energy eigenstates and can be obtained as superpositions of handed states.[39] It follows

$$|\Psi_{\pm}\rangle = \frac{1}{\sqrt{2}}(|\Psi_L\rangle \pm |\Psi_R\rangle) \tag{60}$$

which are, respectively, even- and odd- parity eigenstates. Chirality operators $\chi$ and $\bar{\chi}$ do not alter the parity of a given enantiomer being

$$\chi|\Psi_{\pm}\rangle = \frac{1}{\sqrt{2}}(\chi|\Psi_L\rangle \pm \chi|\Psi_R\rangle) = \frac{1}{\sqrt{2}}(|\Psi_L\rangle \pm |\Psi_R\rangle) \tag{61}$$

$$\bar{\chi}|\Psi_{\pm}\rangle = \frac{1}{\sqrt{2}}(\bar{\chi}|\Psi_L\rangle \pm \bar{\chi}|\Psi_R\rangle) = \frac{1}{\sqrt{2}}(|\Psi_R\rangle \pm |\Psi_L\rangle) \tag{62}$$

which means that $\bar{\chi}$-operators allow transitions between $|\Psi_R\rangle$ and $|\Psi_L\rangle$ (as in a quantum tunnelling process) and parity is the conserved quantum mechanical quantity.[30] It is worth noting that the total Hamiltonian operator for the degenerate isomers of an optically active molecule always consists of an even and an odd part[40]

$$\hat{\mathcal{H}}^{tot} = \hat{\mathcal{H}}^{even} + \hat{\mathcal{H}}^{odd} \tag{63}$$

This is the energy operator involved in the Hund result, which is

$$\left[\hat{P}, \hat{\mathcal{H}}^{tot}\right] = 0 \tag{64}$$

On the other hand, the Hamiltonian operators considered above (i.e. $\hat{\mathcal{H}}_k$) refer to bond-eigenstates of the k-tetrahedron. Due to relations (49) and (50), these eigenstates are "conserved" with respect to the Hamiltonian $\hat{\mathcal{H}}_k$ and not with respect to the total Hamiltonian $\hat{\mathcal{H}}^{tot}$, so then parity and not chirality is the true conserved quantum mechanical quantity.

## 8. SUMMARY AND CONCLUSIONS

At this point, it is interesting to quote Heisenberg's remark[41] which suggested that elementary particles are much more akin to molecules than to atoms. This statement underlines the importance of fundamental symmetry arguments to pursue analogies between quantum states of chiral molecules and those of elementary particles. Hence, the developments in fundamental physics can give access to concepts which could form the basis of a new quantum chemistry. With this perspective in mind, we have developed a new description of chirality of tetrahedral molecules, which takes into account the geometrical and algebraic structure of such objects with implications for their quantum mechanical properties.

On the basis of empirical Fischer projections, it is possible to derive an algebraic approach to central molecular chirality of tetrahedral molecules. The elements of such an algebra are obtained from the 24 projections which a single chiral tetrahedron can generate in $S$ and $R$ configurations. They constitute a matrix representation of $O(4)$ orthogonal group. Twelve of them are rotations, while the other twelve are inversions. All the projections are algebraically generated starting from a fundamental one, where the positions of chemical groups are established *a priori* in a clockwise sequence $1 \rightarrow 2 \rightarrow 3 \rightarrow 4$. The generalization to chains of tetrahedra is straightforward.

According to this representation, given a molecule with $n$ chiral centres, it is possible to define an index of chirality $\chi \equiv \{n, p\}$, where $n$ is the number of stereogenic centres of the molecule and $p$ the number of permutations observed under rotations and superimposition of the tetrahedral molecule to its mirror image.

Consequently, a "chirality selection rule" comes out which allows the characterization of a molecule as *achiral*, *enantiomer* or *diastereoisomer*. The chirality index, not only assigns the global chirality of a given tetrahedral chain, but indicates also a way to predict the same property for new compounds, which can be built up. In fact, a sort of molecular Aufbau for tetrahedra has been proposed. It is possible to recognize a set of rules which allows the classification of new compounds, obtained after the addition of another chiral centre, by the determinantion of the selection rule $\Delta p = 0,1$ with respect to the added centre.

Such a chiral algebra can be discussed in the framework of quantum mechanics. In fact, it is possible to show that the elements of the $O(4)$ group are operators, which commute with the Hamiltonian of the system, and give rise to Heisenberg relations implying conservations laws. In this sense, this algebra can be defined as a "quantum chiral algebra". Moreover, the operators, acting on the molecular chiral states, preserve the parity of the whole system as stated by Hund.[30]

This result is clearly in agreement with the fact that the true stationary states of the systems are the parity ones, while chiral states $|\Psi_R\rangle$ and $|\Psi_L\rangle$ can be interchanged by a quantum tunneling mechanism.[37a]

These new perspectives can give rise to a wide debate on the role of group theory in order to seek for the fundamental features of chirality.